\documentclass{PoS}
\usepackage{graphicx}
\def\mathswitchr#1{\relax\ifmmode{\mathrm{#1}}\else$\mathrm{#1}$\fi}
\newcommand{\PH}{\mathswitchr H}
\newcommand{\Pp}{\mathswitchr p}
\newcommand{\Pt}{\mathswitchr t}
\def\ptcut{{p_{\mathswitchr T}^{\mathswitchr{cut}}}}

\def\mathswitch#1{\relax\ifmmode#1\else$#1$\fi}
\newcommand{\Mt}{\mathswitch {m_\Pt}}

\newcommand{\GeV}{\unskip\,\mathrm{GeV}}
\newcommand{\MeV}{\unskip\,\mathrm{MeV}}

\def\citere#1{\mbox{Ref.~\cite{#1}}}
\def\citeres#1{\mbox{Refs.~\cite{#1}}}

\title{NLO QCD corrections to $\mathswitchr{\bf pp\to t{\bar t} + jet + X}$}

\ShortTitle{NLO QCD corrections to $pp\to{t{\bar t} + jet + X}$}

\author{Stefan Dittmaier$^a$, \speaker{Peter Uwer}$^b$, Stefan Weinzierl$^c$
  \thanks{P.U. is supported as Heisenberg Fellow of the 
    Deutsche Forschungsgemeinschaft DFG. This work is supported
    in part by the European Community's Marie-Curie Research 
    Training Network HEPTOOLS under contract
    MRTN-CT-2006-035505 
    and by the DFG Sonderforschungsbereich/Transregio 9 
    "Computergestützte Theoretische Teilchenphysik" SFB/TR9.}\\
  \llap{$^a$}Max-Planck-Institut f\"ur Physik
  (Werner-Heisenberg-Institut), D-80805 M\"unchen, Germany\\
  \llap{$^b$}Institut fur Theoretische Teilchenphysik, 
         Universitat Karlsruhe, 76128 Karlsruhe, Germany \\
  \llap{$^c$}Universit\"at Mainz, D-55099 Mainz, Germany\\

  E-mail: \email{stefan.dittmaier@mppmu.mpg.de}, %
  \email{uwer@particle.uni-karlsruhe.de},
  \email{stefanw@thep.physik.uni-mainz.de}}

\abstract{%
  We discuss the production of top--anti-top quark pairs in association with
  a hard jet at the Tevatron and at the LHC
  and we report on the calculation of the next-to-leading order QCD
  corrections to this process.
  Numerical results for the $\Pt\bar\Pt$+jet cross
  section and the forward--backward charge asymmetry are presented.
  The corrections stabilize the leading-order prediction for the cross
  section.
  In contrast, the charge asymmetry receives large corrections.
  The dependence of the cross section as well as the asymmetry on the
  minimum transverse momenta used to define the additional jet is
  studied in detail for the Tevatron.
  }

\FullConference{8th International Symposium on Radiative Corrections \\
  October 1-5, 2007\\
  Florence, Italy}

\begin{document}
\section{Introduction}
The top quark is the heaviest of the known elementary particles.
More than ten years after its discovery, the dynamics and
many properties of the top quark,
such as its electroweak quantum numbers, are not yet precisely measured.
It is widely believed that the top quark plays a key role in extensions of
the Standard Model.
This renders experimental investigations of
the top quark particularly important.
Up to now the main (direct) source of information on top quarks are
  top-quark pairs produced at the Tevatron. Only recently 
first evidence for single-top production has been found \cite{Abazov:2006gd}.
It is important to note that in the inclusive $\Pt\bar\Pt$ sample
  a significant fraction comprises $\Pt\bar\Pt$+jet events.
An investigation of the process of $\Pt\bar\Pt$ production in association with
a hard jet can thus improve our knowledge about the top quark.

In this context, the forward--backward charge asymmetry of the
top (or anti-top) quark 
\cite{Halzen:1987xd,Kuhn:1998kw,Kuhn:1998jr,Bowen:2005ap} 
is of particular interest. 
In inclusive $\Pt\bar\Pt$ production it appears first at one loop, because it
results from interferences of C-odd with C-even parts of double-gluon 
exchange between initial and final states. 
This means that the available prediction for $\Pt\bar\Pt$
production---although of one-loop order---describes 
this asymmetry only at leading-order (LO) accuracy.
In $\Pt\bar\Pt{+}$jet production the asymmetry appears already in LO.
Thus, the next-to-leading order (NLO) 
calculation described in the following provides a true
NLO prediction for the asymmetry.
Our calculation will, therefore, be an important tool in the 
experimental analysis of this observable at the Tevatron where 
the asymmetry is measureable as discussed in \citere{Bowen:2005ap}.

Measuring the cross section of the related process
of $\Pt\bar\Pt{+}\gamma$ production provides  direct access 
to the electric charge of the top quark. Obviously NLO QCD predictions
to this process are important for a reliable analysis. 
They can be obtained from $\Pt\bar\Pt{+}$jet production presented here
via simple substitutions.
Finally, a signature of $\Pt\bar\Pt$ in association with a hard jet
represents an important background process for searches at the LHC,
such as the search for the Higgs boson in the weak-vector-boson
fusion or $\Pt\bar\Pt\PH$ channels.

The above-mentioned issues clearly underline the case for an NLO
calculation for $\Pt\bar\Pt{+}$jet production at hadron colliders.
We report here on a first calculation of this kind
as presented in \citere{Dittmaier:2007wz}.

\section{Details of the NLO calculation}
At LO, hadronic $\Pt\bar\Pt{+}$jet production receives contributions
from the partonic processes $q\bar q\to\Pt\bar\Pt g$, $qg\to\Pt\bar\Pt q$, 
$\bar qg\to\Pt\bar\Pt \bar q$, and $gg\to\Pt\bar\Pt g$.
The first three channels are related by crossing symmetry to the 
amplitude
$0 \to \Pt \bar \Pt q \bar q g$.
Evaluating $2\to3$ particle processes at the NLO level, is 
non-trivial, both in the analytical and numerical parts of
the calculation.
In order to prove the correctness of our results we have evaluated 
each ingredient twice using independent calculations based---as
far as possible---on different methods, 
yielding results in mutual agreement.

\subsection{Virtual corrections}
The virtual corrections modify the partonic processes that are
already present at LO. At NLO these corrections
are induced by self-energy, vertex, 
box (4-point), and pentagon (5-point) corrections. The most complicated
diagrams are the pentagon diagrams.

\vspace*{-0.2cm}
\paragraph{Version 1} of the virtual corrections is essentially obtained 
following the method described in \citere{Beenakker:2002nc}, where
$\Pt\bar\Pt\PH$ production at hadron colliders was considered.
Feynman diagrams and amplitudes have been generated with the
{\sl FeynArts} package \cite{Kublbeck:1990xc,Hahn:2000kx}
and further processed with in-house {\sl Mathematica} routines,
which automatically create an output in {\sl Fortran}.
The IR (soft and collinear) singularities are analytically separated
from the finite remainder as described in
\citeres{Beenakker:2002nc,Dittmaier:2003bc}.
The tensor integrals appearing in the pentagon diagrams
are directly reduced to box 
integrals following \citere{Denner:2002ii}. This method does not
introduce inverse Gram determinants in this step, thereby avoiding
notorious numerical instabilities in regions where these determinants
become small. Box and lower-point integrals are reduced 
\`a la Passarino--Veltman \cite{Passarino:1978jh} to scalar integrals,
which are either calculated analytically or using the results of
\citeres{'tHooft:1978xw,Beenakker:1988jr,Denner:1991qq}. 
Sufficient numerical stability is already achieved in this
way. Nevertheless the integral evaluation is currently further refined
by employing the more sophisticated methods described in
\citere{Denner:2005nn} in order to numerically stabilize the tensor
integrals in exceptional phase-space regions.

\vspace*{-0.2cm}
\paragraph{Version 2} of the evaluation of loop diagrams starts
with the generation of diagrams and amplitudes via {\sl QGRAF} 
\cite{Nogueira:1991ex},
which are then further manipulated with {\sl Form}
\cite{Vermaseren:2000nd} and eventually
automatically translated into {\sl C++} code.
The reduction of the the 5-point tensor integrals to scalar
integrals is performed with an extension of the method described in 
\citere{Giele:2004iy}. In this procedure also
inverse Gram determinents of four four-momenta are avoided.
The lower-point tensor integrals are reduced
using an independent implementation of the Passarino--Veltman procedure.
The IR-finite scalar integrals are
evaluated using the {\sl FF} package 
\cite{vanOldenborgh:1990wn,vanOldenborgh:1991yc}.

\subsection{Real corrections}

The matrix elements for the real corrections are given by
$0 \to \Pt \bar \Pt g g g g$, 
$0 \to \Pt \bar \Pt q \bar q g g$,
$0 \to \Pt \bar \Pt q \bar q q' \bar q'$
and
$0 \to \Pt \bar \Pt q \bar q q \bar q$.
The various partonic processes are obtained from these matrix elements 
by all possible crossings of light particles into the initial state.

The evaluation of the real-emission amplitudes is
performed in two independent ways.
Both evaluations employ 
the dipole subtraction formalism 
\cite{Catani:1996vz,Phaf:2001gc,Catani:2002hc}
for the extraction of IR singularities and for their
combination with the virtual corrections. 

\vspace*{-0.2cm}
\paragraph{Version 1} results from a fully automated calculation
based on helicity amplitudes, as
described in \citere{Weinzierl:2005dd}.
Individual helicity amplitudes are computed with the help of
Berends--Giele recurrence relations \cite{Berends:1987me}.
The evaluation of color factors and the generation of subtraction
terms is automated.
For the channel $g g \to \Pt \bar \Pt g g$ a dedicated
soft-insertion routine \cite{Weinzierl:1999yf} is used for the generation 
of the phase space.

\vspace*{-0.2cm}
\paragraph{Version 2} uses for the LO $2 \to 3$ processes
and the $g g \to \Pt \bar \Pt g g$ process optimized code obtained from
a Feynman diagramatic approach.  As in version~1 standard techniques
like color decomposition and the use of helicity amplitudes are
employed. For the $2\to 4$ processes including light quarks, {\sl
  Madgraph} \cite{Stelzer:1994ta} has been used. The subtraction terms
according to \citere{Catani:2002hc} are obtained in a semi-automatized
manner based on an in-house library written in {\sl C++}.

\section{Numerical results}

In the following we consistently use the CTEQ6 
\cite{Pumplin:2002vw,Stump:2003yu}
set of parton distribution functions (PDFs). In detail, we take
CTEQ6L1 PDFs with a 1-loop running $\alpha_{\mathrm{s}}$ in
LO and CTEQ6M PDFs with a 2-loop running $\alpha_{\mathrm{s}}$
in NLO.
The number of active flavours is $N_{\mathrm{F}}=5$, and the
respective QCD parameters are $\Lambda_5^{\mathrm{LO}}=165\MeV$
and $\Lambda_5^{\overline{\mathrm{MS}}}=226\MeV$.
Note that the top-quark loop in the gluon self-energy is
  subtracted at zero momentum. In this scheme the running of 
$\alpha_{\mathrm{s}}$ is generated solely by the contributions of the
light quark and gluon loops. The top-quark mass is
renormalized in the on-shell scheme, as numerical value we take $\Mt=174\GeV$.

We apply the jet algorithm of \citere{Ellis:1993tq}
with $R=1$ for the definition of the tagged hard jet. Unless stated otherwise
we require a transverse momentum of
$p_{\mathrm{T,jet}}>\ptcut = 20\GeV$ for 
the hardest jet.
The outgoing (anti-)top quarks are neither affected
by the jet algorithm nor by the phase-space cut.
Note that the LO prediction and the virtual corrections are not influenced
by the jet algorithm, but the real corrections are.
\looseness-1

\begin{figure}
\centerline{\includegraphics[width=.4\textwidth]{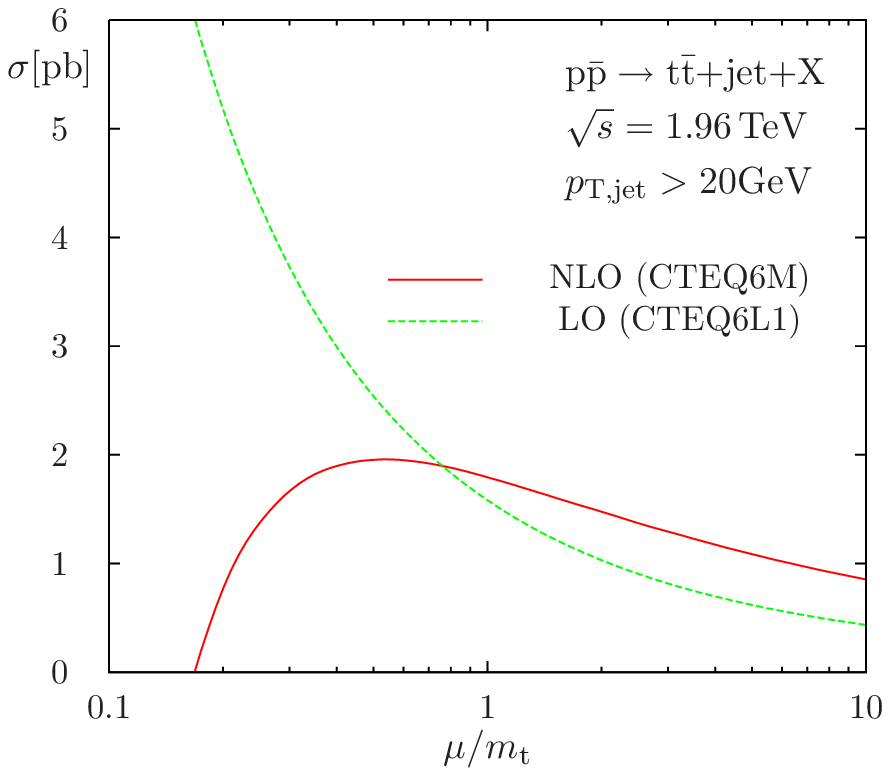}
\includegraphics[width=.4\textwidth]{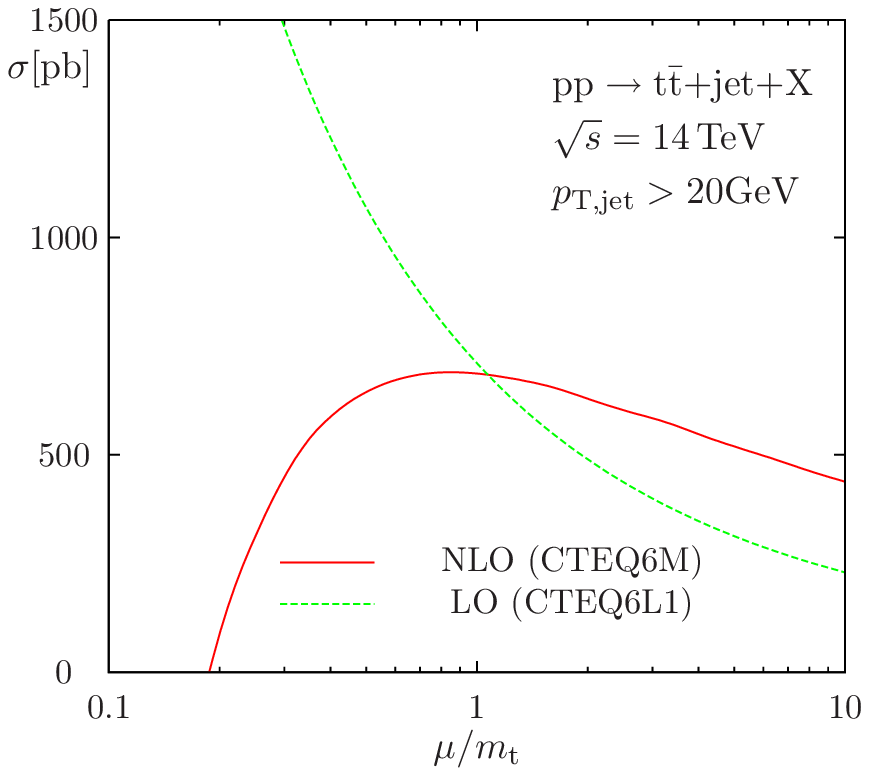}}
\vspace*{-1em}
\caption{Scale dependence of the LO and NLO cross sections for
$\Pt\bar\Pt{+}$jet production at the Tevatron (left) and at the LHC
(right) as taken from \citere{Dittmaier:2007wz}, 
where the renormalization scale ($\mu_r$) and the factorization scale
($\mu_f$) are set
equal to $\mu$.}
\label{fig:NLOcs}
\end{figure}

In Figure~\ref{fig:NLOcs} the scale dependence of the NLO cross sections
is shown. For comparison, the LO results are included as well.  
As expected, the NLO corrections significantly reduce the scale
dependence compared to LO.
We observe that arround $\mu \approx \Mt$ the NLO corrections
are of moderate size for the chosen setup.

We have also studied  the forward--backward 
charge asymmetry of the top quark at the Tevatron.
\begin{figure}
\centerline{\includegraphics[width=.42\textwidth]{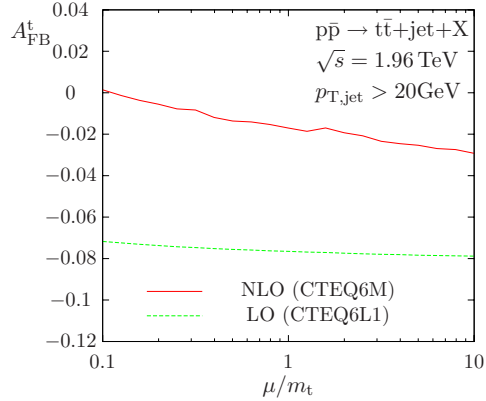}}
\vspace*{-1em}
\caption{Scale dependence of the LO and NLO forward--backward charge asymmetry
of the top quark in $\Pp\bar\Pp\to\Pt\bar\Pt{+}$jet$+X$ at the
Tevatron  as taken from \citere{Dittmaier:2007wz} with $\mu=\mu_f=\mu_r$.}
\label{fig:NLOasym}
\end{figure}
In LO the asymmetry is defined by
\begin{equation}
\label{eq:asym}
 A^{\Pt}_{\mathrm{FB,LO}} = 
\frac{\sigma^-_{\mathrm{LO}}}{\sigma^+_{\mathrm{LO}}},
\quad
\sigma^\pm_{\mathrm{LO}} = 
\sigma_{\mathrm{LO}}(y_{\Pt}{>}0)\pm\sigma_{\mathrm{LO}}(y_{\Pt}{<}0),
\end{equation}
where $y_{\Pt}$ denotes the rapidity of the top quark.
Cross-section contributions 
$\sigma(y_{\Pt}$ \raisebox{-.2em}{$\stackrel{>}{\mbox{\scriptsize$<$}}$} $0)$ 
correspond to top quarks in the forward or backward hemispheres, respectively,
where incoming protons fly into the forward direction by definition.
Denoting the corresponding NLO contributions to the cross sections by
$\delta\sigma^\pm_{\mathrm{NLO}}$,
we define the asymmetry at NLO by
\begin{equation}
\label{eq:NLOasym}
 A^{\Pt}_{\mathrm{FB,NLO}} = 
\frac{\sigma^-_{\mathrm{LO}}}{\sigma^+_{\mathrm{LO}}}
\left( 1+
 \frac{\delta\sigma^-_{\mathrm{NLO}}}{\sigma^-_{\mathrm{LO}}}
-\frac{\delta\sigma^+_{\mathrm{NLO}}}{\sigma^+_{\mathrm{LO}}} \right),
\end{equation}
i.e.\ via a consistent expansion in $\alpha_{\mathrm{s}}$.
Note, however, that the LO cross sections in Eq.~(\ref{eq:NLOasym})
are evaluated in the NLO setup (PDFs, $\alpha_{\mathrm{s}}$).

Figure~\ref{fig:NLOasym} shows the scale dependence of the asymmetry at LO and NLO.
The LO result for the asymmetry is of order $\alpha_s^0$ and does 
therefore not depend on the renormalization scale.
The plot for the LO result shows a mild residual dependence 
on the factorization scale, but the size of this variation does not
reflect the theoretical uncertainty, which is much larger.
The NLO corrections to the asymmetry are of order $\alpha_s^1$ 
and depend on the renormalization scale.
It is therefore natural to expect a stronger scale 
dependence of the asymmetry at NLO than at LO, as seen in the plot.
The size of the asymmetry, which is about $-7\%$ at LO, 
is drastically reduced by the NLO corrections. 
To investigate the origin of the large NLO corrections to the
asymmetry we have studied the dependence on the cut value
$\ptcut$ 
used to define the minimal $p_{\mathswitchr T}$ of the additional
jet. The results are 
shown in Table~\ref{tab:XSection}. We observe that both the
NLO cross section as well as the NLO asymmetry dependent strongly on
$\ptcut$. This is related to the fact that
the cross section becomes ill-defined in the limit $\ptcut \to 0$ due
to the appearance of IR divergencies. On the other hand, the LO
prediction for the asymmetry shows only a mild dependence on 
$\ptcut$. 
\begin{table}[htbp]
  \begin{center}
    \leavevmode
    \begin{tabular}{c|l|l|l|l}
      &\multicolumn{2}{|c|}{cross section [pb]}
      &\multicolumn{2}{|c}{ charge asymmetry [\%]}\\ 
      $\ptcut$ [GeV]&\multicolumn{1}{|c|}{LO}&
      \multicolumn{1}{|c|}{NLO}&
      \multicolumn{1}{|c|}{LO}
      &\multicolumn{1}{|c}{NLO}\\ \hline \hline
      20& 1.583(2)$^{+0.96}_{-0.55}$ & 1.791(1)$^{+0.16}_{-0.31}$ &
      $-7.69(4)^{+0.10}_{-0.085}$ & $-1.77(5)^{+0.58}_{-0.30}$ \\

      30& 0.984(1)$^{+0.60}_{-0.34}$ & 1.1194(8)$^{+0.11}_{-0.20}$ &
      $-8.29(5)^{+0.12}_{-0.085}$ & $-2.27(4)^{+0.31}_{-0.51}$ \\

      40& 0.6632(8)$^{+0.41}_{-0.23}$ & 0.7504(5)$^{+0.072}_{-0.14}$ &
      $-8.72(5)^{+0.13}_{-0.10}$ & $-2.73(4)^{+0.35}_{-0.49}$ \\

      50& 0.4670(6)$^{+0.29}_{-0.17}$ & 0.5244(4)$^{+0.049}_{-0.096}$
      & $-8.96(5)^{+0.14}_{-0.11}$ & $-3.05(4)^{+0.49}_{-0.39}$\\
      \hline \hline
\end{tabular} 
\caption{Cross section and forward-backward charge asymmetry at the
      Tevatron for
      different values of $\ptcut$ used to define the minimal
      transverse momentum $p_{\mathswitchr T}$ of the additional jet
      ($\mu=\mu_f=\mu_r = m_t$).
      The upper and lower indices are the shifts towards $\mu = m_t/2$
      and $\mu = 2 m_t$.}
    \label{tab:XSection}
  \end{center}
\end{table}

\section{Conclusions}

Predictions for $\Pt\bar\Pt{+}$jet production at hadron
colliders have been reviewed at NLO QCD. 
For the cross section the NLO corrections
drastically reduce the scale dependence of the LO predictions, which 
is of the order of 100\%.
The charge asymmtry of the top quarks, which is going to be measured at the
Tevatron, is significantly decreased at NLO and is almost washed out
by the residual scale dependence. In addition we have also studied 
the $\ptcut$-dependence of the NLO predictions. Further refinements
of the precise definition of the forward-backward asymmetry are required
to stabilize the  asymmetry with respect to higher order corrections.



\end{document}